\documentclass{appolb}
\usepackage{epsfig}

\begin{document}
\bibliographystyle{h-elsevier3} 
\title{Hydrodynamics for relativistic heavy ion collisions%
\thanks{Presented by PB at \textit{Excited QCD 2012}, Peniche, Portugal, 6--12 May 2012.}%
}
\author{
P.~Bo\.zek$^{1, 2}$, W.~Broniowski$^{1,3}$,
I.~Wyskiel-Piekarska$^1$
\address{
$^1$Institute of Nuclear Physics
    PAN, PL-31342 Krak\'ow, Poland\\
$^2$Institute of Physics, Rzesz\'ow University, 
PL-35959 Rzesz\'ow, Poland\\
$^3$Institute of Physics, Jan Kochanowski University, PL-25406~Kielce, Poland}
}
\maketitle
\begin{abstract}
Simulations of the  viscous 
hydrodynamic model for relativistic heavy-ion collisions at RHIC and LHC energies are presented.
Results for spectra,  femtoscopy radii, and transverse momentum fluctuations are favorably compared to the experimental data.
Effects of the local charge conservation on correlation observables are also studied.
\end{abstract}
  
\section{Introduction}

The expansion of the fireball in heavy-ion collisions generates collective, azimuthally asymmetric transverse flow which can be described
quantitatively in hydrodynamic calculations 
\cite{Kolb:2003dz,Florkowski:2010zz,Luzum:2008cw,Song:2010aq,Bozek:2009dw,Schenke:2010rr,Broniowski:2008vp}. 
Due to large velocity gradients in the system, an important role in the 
 hydrodynamic model is played by shear and bulk viscosities
 \cite{IS,Romatschke:2009im,Teaney:2009qa}. 
 The asymmetry of the final flow is largely determined by  fluctuations of the initial geometry of the fireball
\cite{Alver:2008zz,Alver:2010gr}. We apply a $3+1$-dimensional viscous hydrodynamic model \cite{Bozek:2011ua} 
 to Pb-Pb collisions at $2.76$~TeV using averaged initial conditions, and to Au-Au collisions at $200$~GeV using fluctuating 
initial conditions. Also, to estimate  the collective flow in 
p-Pb and d-Pb interactions at the LHC,  event-by-event simulations  are used
 \cite{Bozek:2011if}. The angular dependence in the 
two-dimensional dihadron correlation functions for particles with soft momenta are 
determined by the collective flow \cite{Takahashi:2009na}. The short-range charge dependent structures in the same side ridge 
can be  explained  largely as an effect of the local charge conservation at hadronization \cite{Bozek:2012en}.

Initial conditions must be provided for the hydrodynamic evolution. No first-principle calculation of the
formation and thermalization of the dense matter in the early stage exists yet. In our simulations we use the optical 
Glauber model for the averaged initial conditions, and
its Monte Carlo implementation \cite{Broniowski:2007nz} for the fluctuating initial conditions.
The parameters of the model are adjusted to the RHIC  data 
\cite{Bozek:2009dw},  with the default values of the shear and bulk viscosity coefficients $\eta/s=0.08$ and $\zeta/s=0.04$ 
(bulk viscosity is present only in the hadronic phase).
The emission of particles and resonance decays at  freeze-out
are performed via the event generator THERMINATOR \cite{Kisiel:2005hn}.

\section{Averaged initial conditions}

\begin{figure}[b]
\begin{center}
\includegraphics[angle=0,width=0.51\textwidth]{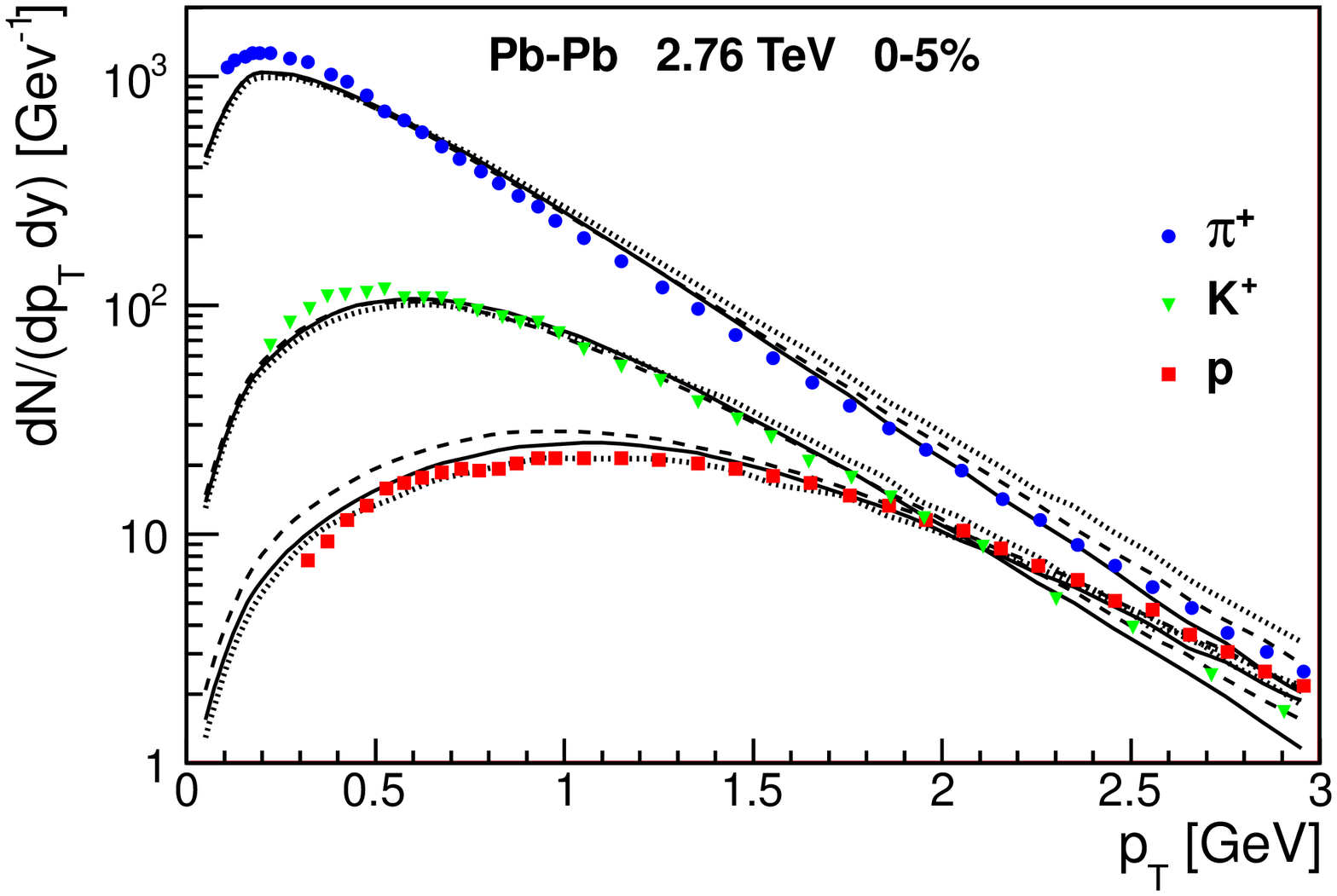}
\includegraphics[angle=0,width=0.48\textwidth]{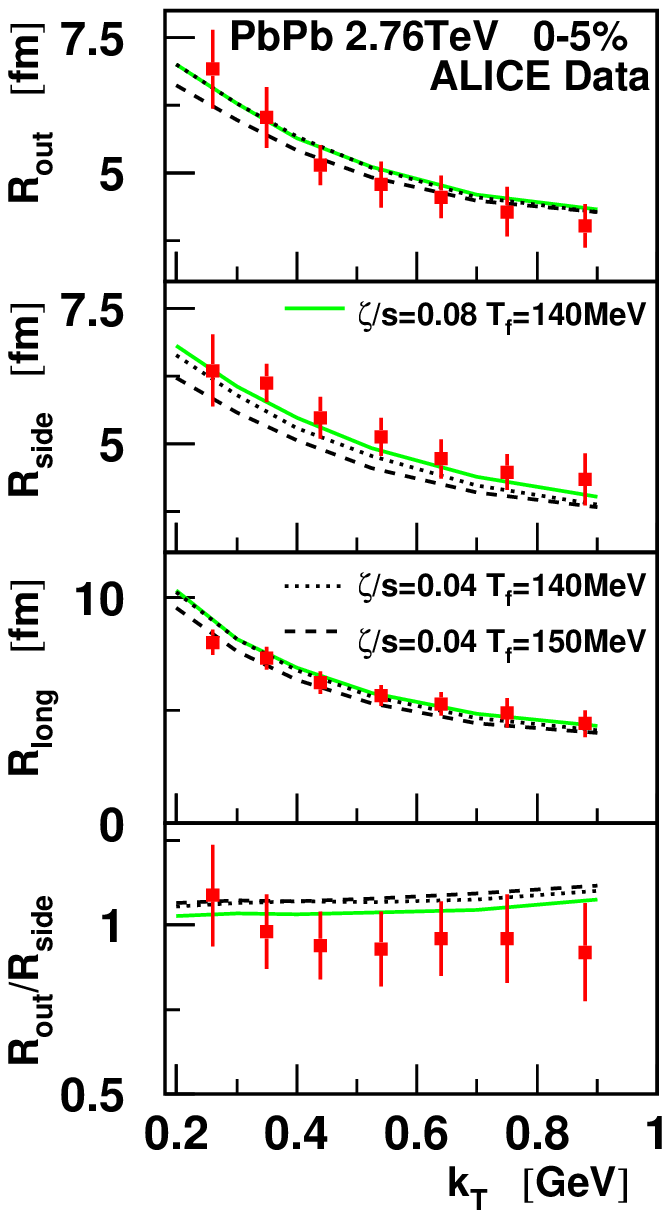}
\end{center}
\vspace{-2mm}
\caption{\small (left) Preliminary data on 
$\pi^+$, $K^+$, $p$ transverse
momentum spectra 
   \cite{Floris:2011ru} and (right) 
the femtoscopic radii  measured by the ALICE 
Collaboration \cite{Aamodt:2011mr}, 
 compared to
 viscous hydrodynamic results  (from \cite{Bozek:2012qs}).}
\label{fig:spectra}
\end{figure}

The use of the averaged conditions in  $3+1$-dimensional viscous hydrodynamics leads to a satisfactory description
of pseudorapidity distributions, transverse momentum spectra, elliptic flow, and the femtoscopic radii in Au-Au interaction at the top 
RHIC energies \cite{Bozek:2011ua}. Hydrodynamic expansion with 
a hard equation of state with cross-over generates a strong transverse flow. It is interesting to note  
that the shear viscosity leads to a reduction of the odd component of the directed flow. 

In Pb-Pb collisions at $\sqrt{s}=2.76$~TeV the transverse collective flow is even 
stronger. One observes a significant shift of the spectra of protons towards higher $p_T$, consistent with 
the preliminary ALICE data (Fig.~\ref{fig:spectra}). 
On the other hand,
 the calculated spectra of pions and kaons are too flat. A slightly better description is obtained with 
a larger value of the bulk viscosity, $\zeta/s=0.08$.
We find that an initial energy density compatible with the observed charged particle pseudorapidity distribution leads
to approximate plateau in the rapidity distributions of identified particles. 
The model reproduces very well the femtoscopic radii, which indicates that the basic features of the space-time evolution 
of the system are realistic.

\section{Fluctuating initial conditions}

\begin{figure}[t]
\includegraphics[angle=0,width=0.37\textwidth]{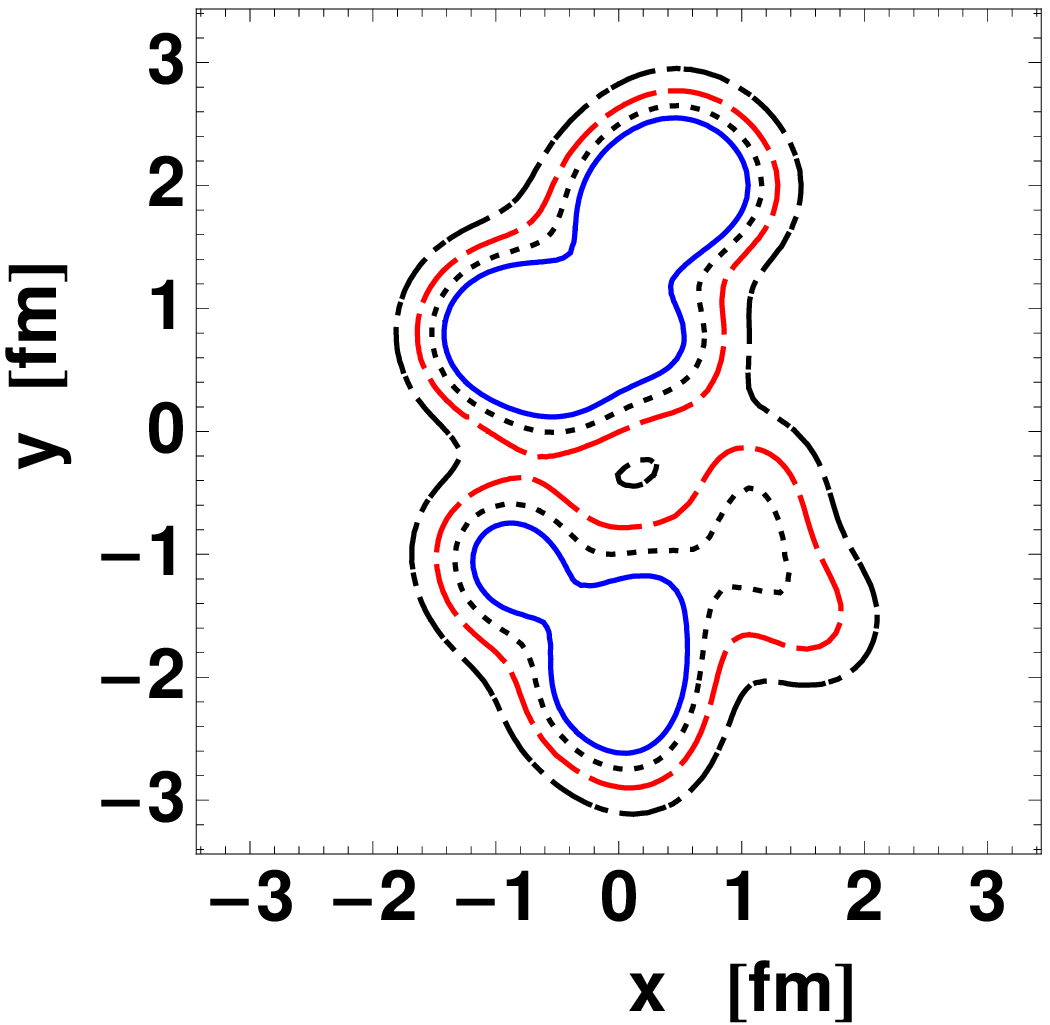}~~~~
\includegraphics[angle=0,width=0.52\textwidth]{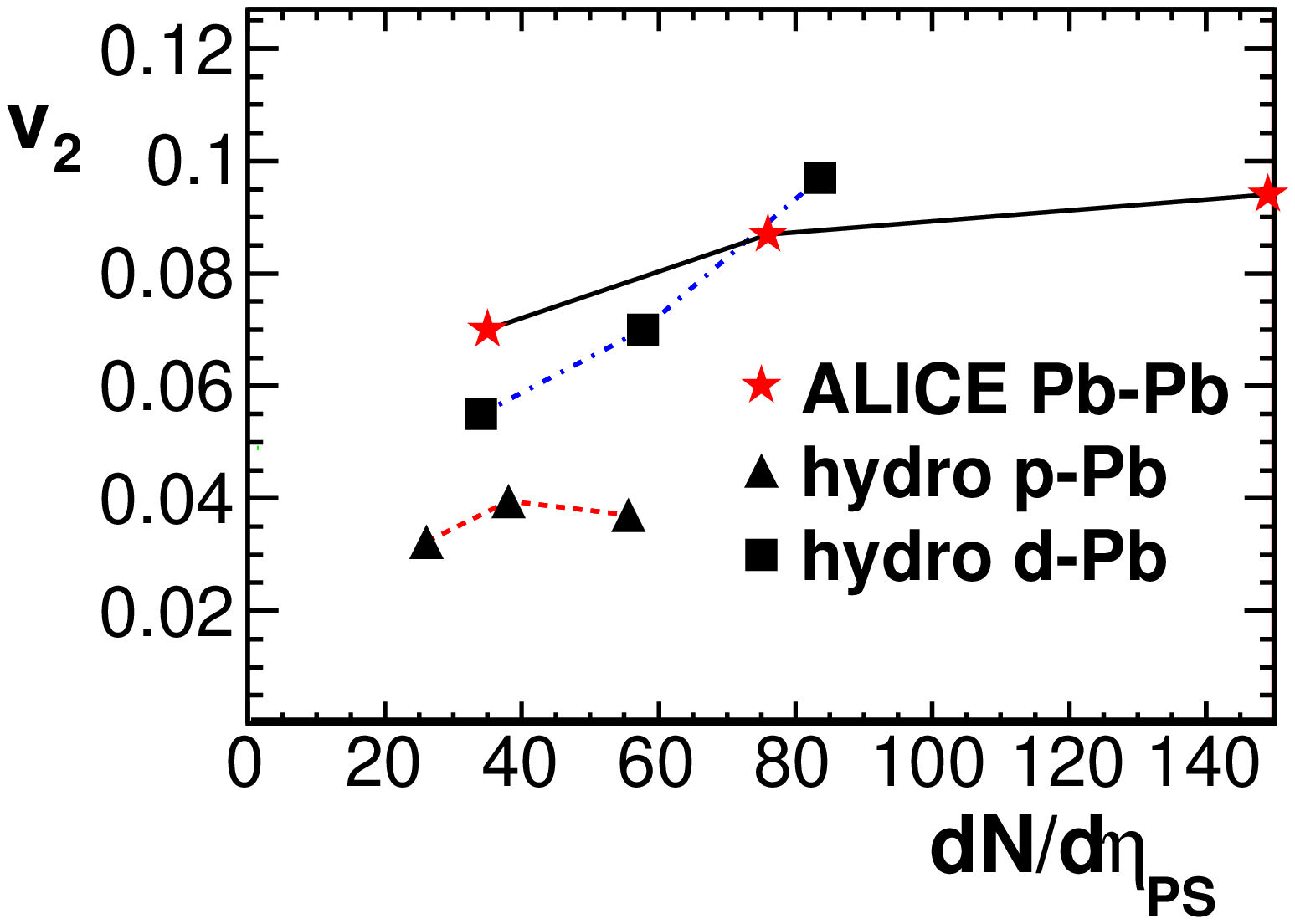}
\caption{\small (left) Initial entropy density in the transverse plane in a d-Pb collision. (right)
 Elliptic flow coefficient predicted in p-Pb collisions at $4.4$~TeV and d-Pb collisions at $3.11$~TeV (from \cite{Bozek:2011if}). }
\label{fig:ppb}
\end{figure}

Event by event 
fluctuations in the initial conditions induce additional  deformations of the  transverse profile of the fireball.
The flow coefficients of the elliptic ($v_2$) and triangular ($v_3$) flow are determined by the initial eccentricity 
and triangularity of the fireball \cite{Alver:2008zz,Alver:2010gr,Schenke:2010rr}. Relatively, the most violent fluctuations 
occur in peripheral collisions or in small systems. For p-Pb and d-Pb collisions at the LHC energies the fireball is large 
enough to allow for a noticeable stage of collective expansion. The initial deformation of the density profile is large, 
especially in d-Pb interactions (Fig.~\ref{fig:ppb}). The generated elliptic and triangular flows are comparable
to values observed in peripheral Pb-Pb collisions and could be measured 
in future experiments \cite{Bozek:2011if}. If such a scenario is realized, the spectra measured in p-Pb and d-Pb collisions 
should be strongly modified by final state (hydrodynamic) interactions, which means that such observations
 could not be used as reference data when looking for  medium modifications expected in Pb-Pb collisions. 
 
\begin{figure}[t]
\includegraphics[angle=0,width=0.495\textwidth]{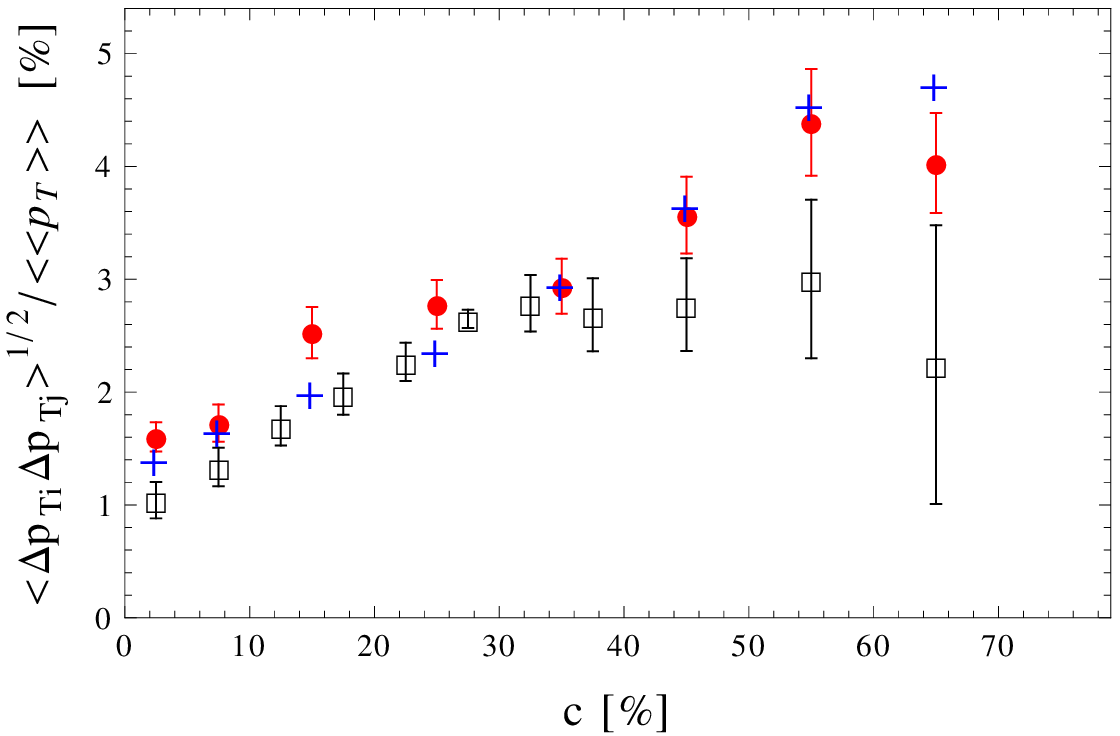}
\includegraphics[angle=0,width=0.495\textwidth]{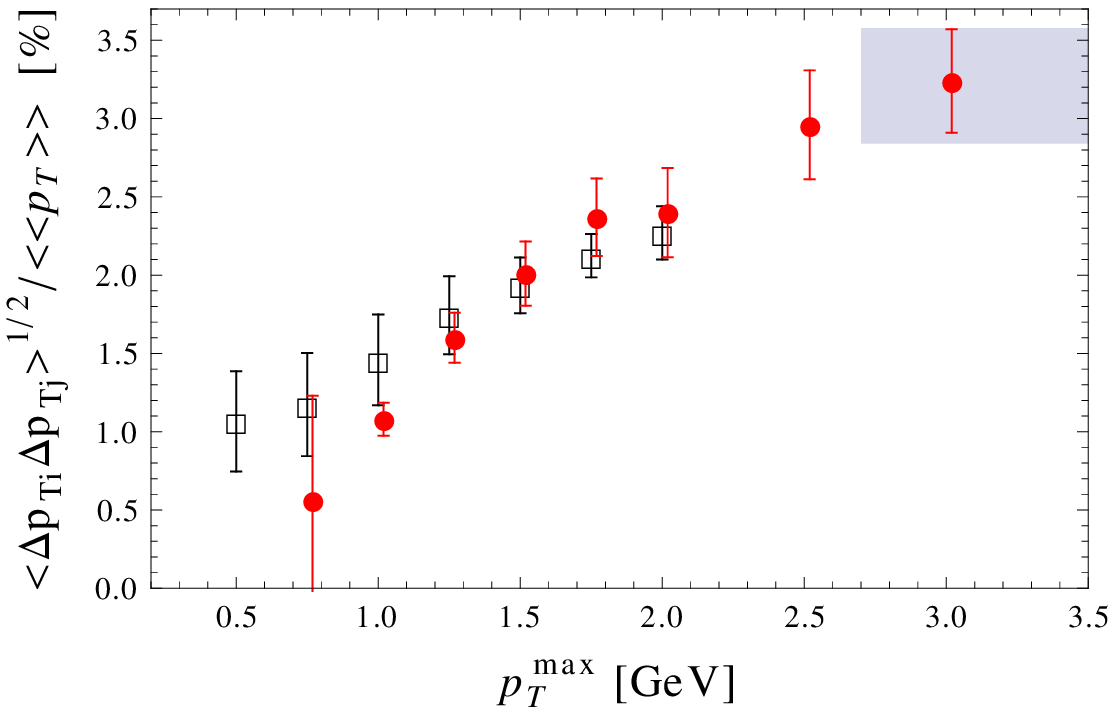}
\caption{\small  The experimental data for
 $\langle \Delta p_{Ti} \Delta p_{Tj} \rangle^{1/2}/\langle\langle p_T  \rangle\rangle$ 
 from the PHENIX Collaboration
\cite{Adler:2003xq} (squares) compared to simulations with event-by-event
viscous hydrodynamics (dots) and to  the approximate result from Ref.~\cite{Broniowski:2009fm}
 for perfect hydrodynamics with smooth  conditions (crosses) (from \cite{Bozek:2012fw}).}
\label{fig:ptfl}
\end{figure}

In Au-Au collisions at $200$~GeV our calculations give  elliptic and triangular flows  similar as in other calculations
\cite{Schenke:2010rr,Alver:2010dn,Petersen:2010cw,Qiu:2011fi}, 
showing a reduction of the flow asymmetry with an increasing shear viscosity.
An additional effect of the hydrodynamic response to fluctuations is seen in the fluctuations of the average momentum 
in an event  \cite{Broniowski:2009fm,Bozek:2012fw}. Both the shape and the size of the  initial fireball
fluctuates event by event.  While the role of the shape fluctuations in generating azimuthally asymmetric flow is widely 
discussed, the size fluctuations of the fireball are less studied. Our simulations show that
an anti-correlation is seen
between the r.m.s radius of the fireball and the average transverse momentum of the emitted particles.
The experimental measure of the transverse momentum fluctuations eliminates statistical fluctuation due the finite 
multiplicity in the events. The remaining scaled transverse momentum fluctuations 
are proportional to the scaled fluctuations of the fireball size
${\Delta p_T}/{\langle p_T \rangle } \simeq 0.3 {\Delta r}/{\langle r \rangle}$.
The size fluctuations at each centrality are predicted in the Glauber model.
The calculated $p_T$ fluctuations are similar as in the data, although some overestimate of the fluctuations can be seen
(Fig.~\ref{fig:ptfl}). We have verified that the hydrodynamic response  is not noticeably modified by
a change in the freeze-out temperature, the bulk and shear viscosities, or the core-corona effects. 
The scaled transverse momentum fluctuations grow with increasing upper momentum cut-off used in the analysis. The same 
behavior in seen in our calculation, where it originates from the hydrodynamic response to the size fluctuations.

 \section{Non-flow charge correlations}

The assumption that charge formation happens  late in the evolution (at hadronization) explains the observed  charge 
balance functions in relative pseudorapidity \cite{Bass:2000az,Jeon:2001ue,Bozek:2003qi,Aggarwal:2010ya}. A similar effect 
is expected and observed in the relative
azimuthal angle of the emitted particles \cite{Bozek:2004dt,Aggarwal:2010ya}. By generalizing these arguments to the 
two-dimensional dihadron correlation function in $\Delta \phi-\Delta \eta$, we expect the appearance of a two-dimensional
same-side  peak for the unlike-sign pairs.

\begin{figure}[h]
\includegraphics[height=4cm]{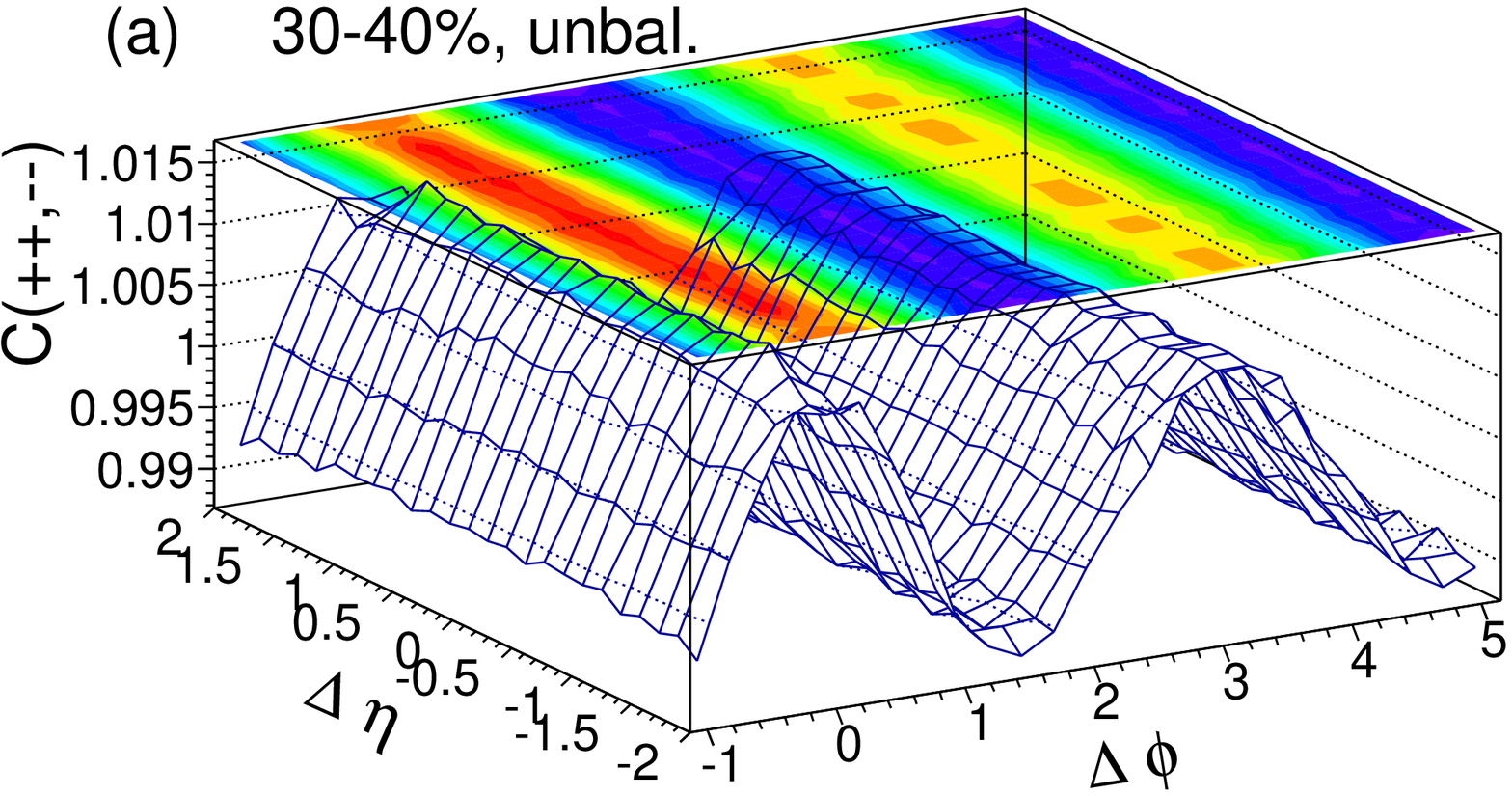}~
\includegraphics[height=4cm]{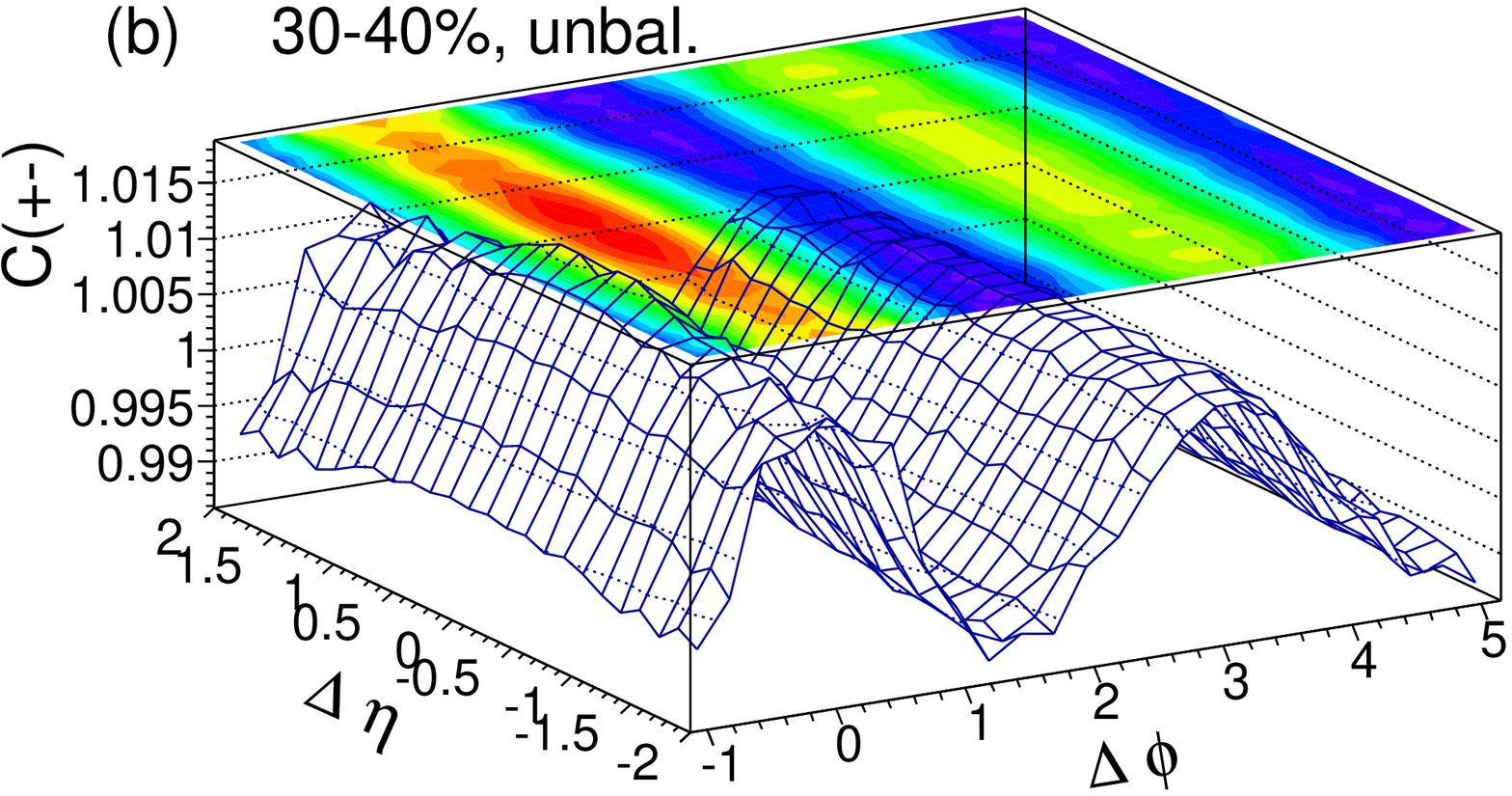} \vspace{-7mm} \\
\includegraphics[height=4cm]{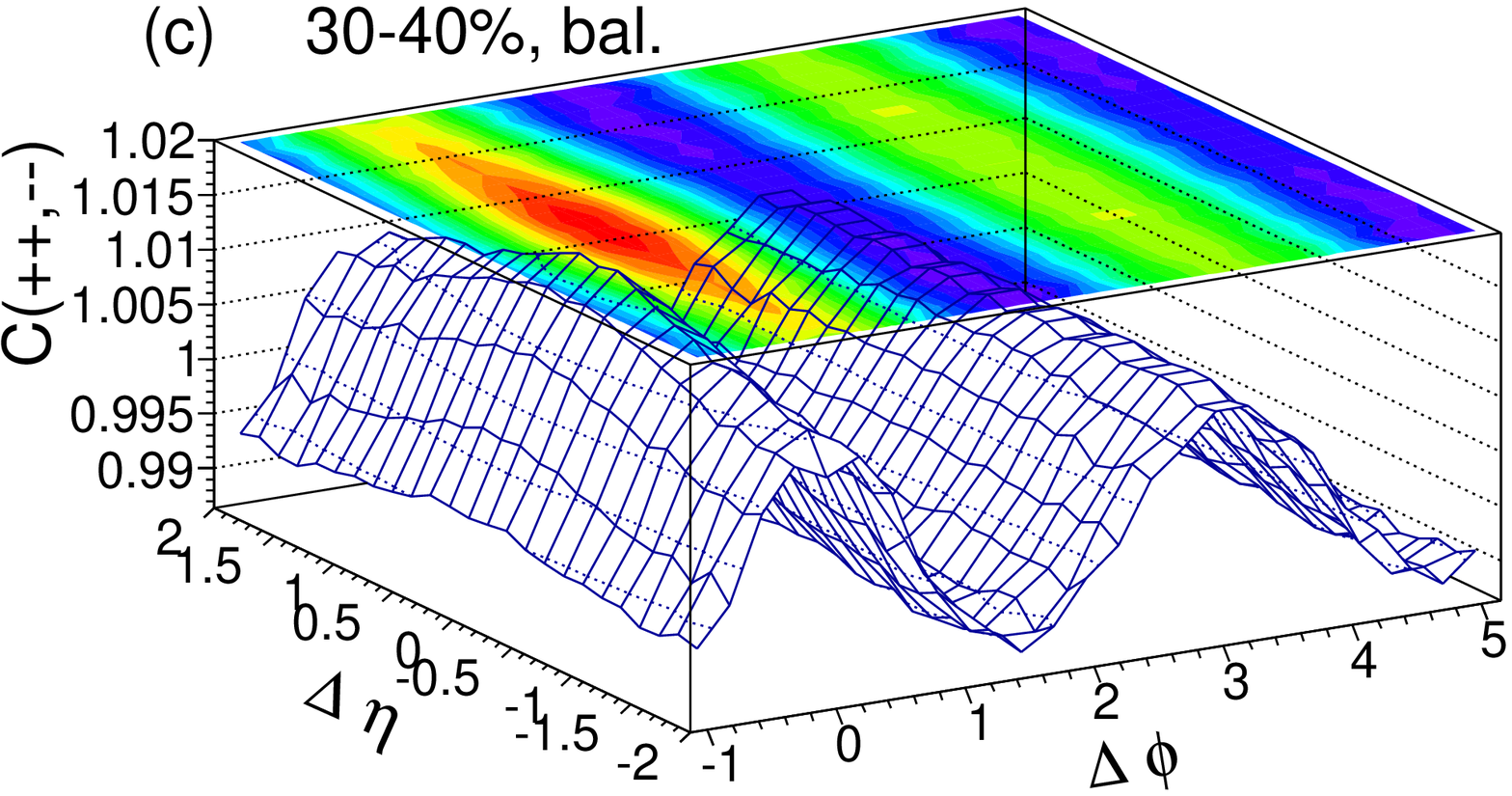}~
\includegraphics[height=4cm]{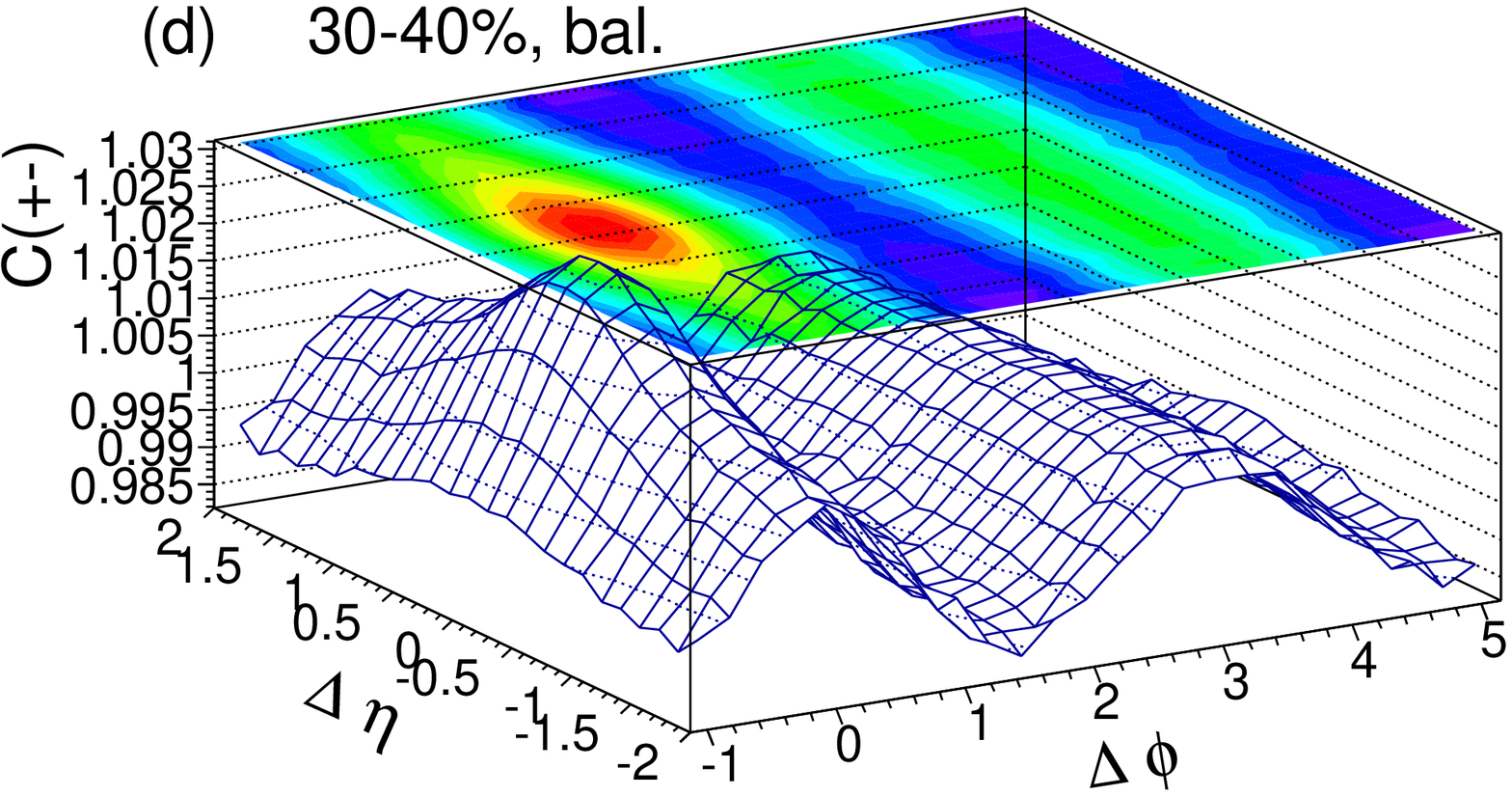} \vspace{-2mm}
\caption{\small  Dihadron correlation functions  for unlike- and like-sign pairs without (upper row) or with (lower row)
 charge balancing mechanism implemented at the end of the hydrodynamic evolution (from \cite{Bozek:2012en}).}
\label{fig:dih}
\end{figure}

 After adding such a charge balancing effect 
to the particle emission at freeze-out \cite{Bozek:2012en},
 qualitatively the same correlations are formed as 
observed experimentally \cite{Agakishiev:2011pe}  (Fig. \ref{fig:dih}). This non-flow  effect is added to the dominant 
 structure from the collective flow in the  correlation functions. For the like-sign pairs 
the same-side ridge is nearly flat, showing essentially no additional structure. 
The charge-balancing also leads to non-flow correlations, increasing the observed flow coefficients $v_2$ and $v_3$.

\vskip 3mm

Supported by
Polish Ministry of Science and Higher
Education, grant N~N202~263438 and N~N202~086140, and National Science
Centre, grant DEC-2011/01/D/ST2/00772.

\bibliography{../../hydr}
\end{document}